\documentclass[amsmath,amssymb,showpacs,draft,preprint,floatfix]{revtex4}
\usepackage{graphicx}
\usepackage{latexsym}
\begin{document}
\title{\bf Direct measurement of the $Q$-function in a lossy cavity}
\author{H. Moya-Cessa${}^1$, R. Ju\'arez-Amaro${}^{1,2}$, and I. Ricardez-Vargas${}^1$}
\affiliation{$^1$Instituto Nacional de Astrof\'{\i}sica, Optica y
Electr\'onica, Apdo. Postal 51 y 216, 72000 Puebla, Pue., Mexico
\\ $^2$Universidad Tecnol\'ogica de la Mixteca, Apdo. Postal 71,
69000 Huajuapan de Le\'on, Oax., Mexico}
\date{\today}
\begin{abstract}
We show that the $Q$-function corresponding to an electromagnetic
field in a lossy cavity can be directly measured by means of a
simple scheme, therefore allowing the knowledge of the state of
the field despite dissipation.
\end{abstract}
 \maketitle

The reconstruction of a quantum state is a central topic in
quantum optics and related fields \cite{ris,leo}. During the past
years, several techniques have been developed, for instance the
direct sampling of the density matrix of a signal mode in
multiport optical homodyne tomography \cite{zuk}, tomographic
reconstruction by unbalanced homodyning \cite{wal}, reconstruction
via photocounting \cite{ban}, cascaded homodyning \cite{kis} to
cite some. There have also been proposals to measure
electromagnetic fields inside cavities \cite{lut,moy} and
vibrational states in ion traps \cite{lut,bar}. In fact the full
reconstruction of nonclassical states of the electromagnetic field
\cite{smi} and of (motional) states of an ion \cite{lei} have been
experimentally accomplished. The quantum state reconstruction in
cavities is usually achieved through a finite set of selective
measurements of atomic states \cite{lut} that make it possible to
construct quasiprobability distribution functions such as the
Wigner function, that constitute an alternative representation of
a quantum state of the field.

Nevertheless, in real experiments, the presence of noise and
dissipation has normally destructive effects. Schemes that treat a
lossy cavity have been  proposed \cite{moy} that involve a
physical process that allows the storage of information about the
quantum coherences of the initial state in the diagonal elements
of the density matrix of a transformed state. Here, we would like
to re-take the problem of  reconstruction of the cavity field as
studied in \cite{lut}, but allowing the cavity to have losses
(i.e. a real cavity). We then will show, that although it is not
possible to reconstruct the Wigner function, it is still possible
to recover whole information about the initial state via the
$Q$-function.

Let us first consider the ideal case of no dissipation. We
consider then the Hamiltonian  for the interaction between a
quantized field and a two-level atom reads (we have set $\hbar=1$)

\begin{equation}
\hat{H}= \omega \hat{a}^{\dagger} \hat{a} +
\frac{\omega_{eg}}{2}\hat{\sigma}_z +
\lambda(\hat{a}^{\dagger}\hat{\sigma}_- +\hat{a}\hat{\sigma}_+ ),
\end{equation}
where $\hat{a}^{\dagger}$ and $\hat{a}$ are the creation and annihilation operators
for the field mode, respectively, obeying $[\hat{a},\hat{a}^{\dagger}]=1$.
$\hat{\sigma}_+=|e\rangle\langle g|$ and $\hat{\sigma}_-=|g\rangle\langle e|$
are the raising and lowering
atomic operators, respectively, $|e\rangle$ being the excited state
 and $|g\rangle$ the ground state of the two-level atom.
The atomic operators obey the commutation relation
$[\hat{\sigma}_+,\hat{\sigma}_-]=\hat{\sigma}_z$. $\omega$
is the field frequency, $\omega_{eg}$ the atomic frequency and $\lambda$ is the
interaction constant.
When we have the condition on the detuning, $\delta=\omega_{eg}-\omega$,
\begin{equation}
\frac{|\delta|}{\lambda} \gg \sqrt{n+1}
\end{equation}
for any "relevant" photon number,
we can obtain an effective interaction Hamiltonian
in the dispersive limit (see for instance \cite{pei})

\begin{equation}
\hat{H}_I^{eff} = \chi \hat{a}^{\dagger} \hat{a} \hat{\sigma}_z,
\label{dis}
\end{equation}
with $\chi= \lambda^2/\delta$.

 Let us consider the atom initially
in the following superposition
\begin{equation}
|\psi_A(0)\rangle=\frac{1}{\sqrt{2}}(|e\rangle+|g\rangle)
\end{equation}
and the state of the field to be arbitrary, denoted by
$|\psi_F(0)\rangle$. If before the interaction we displace the
unknown field by a quantity $\alpha$, we have the state for the
total initial wave function given by
\begin{equation}
|\psi(\alpha;0)\rangle=\frac{1}{\sqrt{2}}(|e\rangle+|g\rangle)|\psi_F(\alpha;0)\rangle
, \label{ini}
\end{equation}
where $|\psi_F(\alpha;0)\rangle=\hat{D}(\alpha)|\psi_F(0)\rangle$
with $\hat{D}(\alpha)=\exp(\alpha\hat{a}^{\dagger}
-\alpha^*\hat{a} )$ is the displacement operator with amplitude
$\alpha$. By calculating the average value of the electric dipole
$\hat{\sigma}_x= \hat{\sigma}_-+\hat{\sigma}_+$, we obtain
\cite{lut,expl}
\begin{equation}
\langle \hat{\sigma}_x\rangle =
\sum_{m=0}^{\infty}P_m(\alpha;0)\cos(2\chi m t) \label{wig}
\end{equation}
where  the photon distribution $P_m(\alpha;0)= |\langle
\psi_F(\alpha;0)|m \rangle|^2$ with $|m \rangle$ a number state.
By choosing $t=\pi/(2\chi)$, we end up with
\begin{equation}
\langle \hat{\sigma}_x\rangle =
\sum_{m=0}^{\infty}P_m(\alpha;0)(-1)^m
\end{equation}
that, except for a factor of $\pi$ is the Wigner function \cite{lut,kni}.

The above treatment does not consider dissipation, and it is the
aim of this manuscript to study the case when we have a
dissipative cavity. The problem of the interaction of a two-level
atom  with a quantized field in the dispersive regime in a cavity
with losses was treated exactly by Peixoto and Nemes \cite{pei}.
Here we will use a superoperator technique to solve it in an
alternative way.

In the interaction picture, and in the dispersive approximation, the master equation that governs
the dynamics of a two-level atom coupled with an electromagnetic field in a high-$Q$ cavity is

\begin{equation}
\frac{d}{dt}\hat{\rho}=-i[\hat{H}_I^{eff},\hat{\rho}]+\hat{\cal{L}}\hat{\rho}
\label{mas}
\end{equation}
where
\begin{equation}
\hat{\cal{L}}\hat{\rho} = 2\gamma \hat{a} \hat{\rho} \hat{a}^{\dagger}-\gamma  \hat{a}^{\dagger} \hat{a}\hat{\rho}
-\gamma\hat{\rho}   \hat{a}^{\dagger} \hat{a},
\end{equation}
and $\hat{\rho}$ the density matrix of the system.

We define the superoperators

\begin{equation}
\hat{L}\hat{\rho} = -\hat{\Gamma} \hat{a}^{\dagger} \hat{a}\hat{\rho}
-\hat{\rho} \hat{\Gamma}^{\dagger}  \hat{a}^{\dagger} \hat{a},
\end{equation}
and
\begin{equation}
\hat{J}\hat{\rho} = 2\gamma \hat{a} \hat{\rho} \hat{a}^{\dagger}
\end{equation}
where we have defined
\begin{equation}
\hat{\Gamma}=\gamma\hat{1}_A+i\chi\hat{\sigma}_z,
\end{equation}
with $\hat{1}_A = |e\rangle\langle e|+|g\rangle\langle g|$.
It is not difficult to show that
\begin{equation}
[\hat{J},\hat{L}]\hat{\rho} = - \hat{S}_{\Gamma}\hat{J}\hat{\rho},
\end{equation}
where the superoperator $\hat{S}_{\Gamma}$ is defined as
\begin{equation}
 \hat{S}_{\Gamma}\hat{\rho} = \hat{\Gamma} \hat{\rho} + \hat{\rho} \hat{\Gamma}^{\dagger}.
\end{equation}
The solution to Eq. (\ref{mas}) subject to the initial state
(\ref{ini}) is then given by \begin{equation} \hat{\rho}(t) =
e^{(\hat{L}+\hat{J})t}\hat{\rho}(\alpha;0)
=e^{\hat{L}t}e^{\hat{f}(t)\hat{J}}\hat{\rho}(\alpha;0)
\label{solu}
\end{equation}
where
\begin{equation}
\hat{f}(t)\hat{\rho}=\frac{1-e^{-\hat{S}_{\Gamma}t}}{\hat{S}_{\Gamma}}\hat{\rho}.
\end{equation}
and $\hat{\rho}(\alpha;0)= |\psi(\alpha;0)\rangle
\langle\psi(\alpha;0)|$.

We need to operate the density matrix with the exponential of superoperators
given above. It is not obvious how $e^{\hat{f}(t)\hat{J}}$ will apply on
(\ref{ini}), and therefore we give an expression for it
\begin{equation}
e^{\hat{f}(t)\hat{J}}\hat{\rho}(0;\alpha)=\sum_{n=0}^{\infty}
\frac{\hat{J}^n[\hat{D}(\alpha)|\psi_F(0)\rangle\langle\psi_F(0)|\hat{D}^{\dagger}(\alpha)]
\times \hat{f}^n [|\psi_A\rangle\langle\psi_A|]}{n!}
\label{suma}
\end{equation}
This is, because  $\hat{f}$ is an atomic superoperator, it will
operate only on atomic states, and $\hat{J}$ will operate only on
field states. It is not difficult to show that
\begin{equation}
2\hat{f}^n |\psi_A\rangle\langle\psi_A| =
\frac{(1-e^{-(\xi+\xi^*)t})^n}{(\xi+\xi^*)^n} \times \hat{1}_A
 +  \frac{(1-e^{-2\xi t})^n}{(2\xi)^n} |e\rangle\langle g| +
\frac{(1-e^{-2\xi^*t})^n}{(2\xi^*)^n} |g\rangle\langle e| ,
\label{aplf}
\end{equation}
with $\xi=\gamma+i\chi$.
From (\ref{aplf}), (\ref{suma}) and (\ref{solu}) we calculate $\langle \hat{\sigma}_x\rangle$ and obtain
\begin{eqnarray}
\langle \hat{\sigma}_x\rangle   & = & \frac{1}{2}\sum_{m=0}^{\infty}
\frac{\left(\gamma\frac{1-e^{-2\xi t}}{\xi}\right)^m}{m!}
\sum_{k=0}^{\infty}
e^{-2k\xi t}|\langle k|\hat{a}^m \hat{D}(\alpha)|\psi_F(0)\rangle|^2 + c.c. \nonumber \\
& = &
\frac{1}{2}\sum_{m=0}^{\infty}
\frac{\left(\gamma\frac{1-e^{-2\xi t}}{\xi}\right)^m}{m!}
\sum_{k=0}^{\infty}
e^{-2k\xi t}\frac{(m+k)!}{k!}\langle k + m |\hat{\rho}(0;\alpha)|k+m\rangle + c.c.
\end{eqnarray}
By changing the summation index in the second sum of the above equation, with $M=m+k$,
we obtain
\begin{equation}
\langle \hat{\sigma}_x\rangle   =
\frac{1}{2}\sum_{m=0}^{\infty}
\frac{\left(\gamma\frac{e^{2\xi t}-1}{\xi}\right)^m}{m!}
\sum_{M=m}^{\infty}
e^{-2M\xi t}\frac{(M)!}{(M-m)!}\langle M |\hat{\rho}(0;\alpha)|M\rangle + c.c.
\label{dob}
\end{equation}
Finally, we can start the second sum of (\ref{dob}) from $M=0$ (as
we would only add zeros to the sum, because the factorial of a
negative integer is infinite), and exchange the double sum  in it,
to sum first over $m$, which gives

\begin{equation}
\langle \hat{\sigma}_x\rangle  =
\frac{1}{2}\sum_{M=0}^{\infty}
\left( \frac{\gamma + i \chi e^{-2\xi t}}{\xi}\right)^M
\langle M |\hat{\rho}(0;\alpha)|M\rangle + c.c.
\end{equation}
By defining
\begin{equation}
\theta = \tan^{-1}\left(- \frac{\eta + e^{-2\eta \tau}[\sin (2\tau) -
\eta \cos (2\tau)]}
{\eta^2+e^{-2\eta \tau}[\cos (2\tau) + \eta \sin (2\tau)]}\right)
\end{equation}
and
\begin{equation}
\mu = \left(\frac{\eta^2+ e^{-4\eta \tau} +
2\eta e^{-2\eta \tau} \sin (2 \tau)}
{1+\eta^2}\right)^{\frac{1}{2}}
\label{mu}
\end{equation}
with $\tau=\chi t$ and $\eta=\gamma/\chi$, we can have a final expression for $\langle \hat{\sigma}_x\rangle $
\begin{equation}
\langle \hat{\sigma}_x\rangle   = \sum_{M=0}^{\infty}
\mu^M \cos(M\theta)
\langle M |\hat{\rho}(0;\alpha)|M\rangle  .
\label{fin}
\end{equation}

Equation (\ref{fin}) in general differs from an $s$-parametrized
quasiprobability distribution (see for instance \cite{kni})
because  for $\eta\neq 0$, i.e. the dissipative case, we will have
$\theta \neq \pi$ and therefore $\cos(M\theta)\neq\epsilon^M$,
where $\epsilon$ is a parameter.
However, if we consider the case
when $\mu=0$, the only term that survives in (\ref{fin}) is $M=0$,
and we obtain the $Q$-function.
\begin{equation}
\langle \hat{\sigma}_x\rangle   =  \langle 0
|\hat{\rho}(0;\alpha)|0\rangle  =  \langle \alpha
|\hat{\rho}(0)|\alpha\rangle  = \pi Q(\alpha). \label{final}
\end{equation}

To look for the value that makes $\mu$ equal to zero, we re-write,
(\ref{mu}) for $\tau = 3\pi/4$ in the form
\begin{equation}
\mu(\tau = \frac{3\pi}{4}) = \frac{|\eta - e^{-3\eta\pi/2}|}{(1+\eta^2)^{1/2}},
\end{equation}
and therefore, for $\eta=\exp(-3\eta \pi/2)$ we obtain $\mu=0$.
The value for $\eta$ may be easily obtained numerically with the
result $\eta_{\mu=0} \approx 0.274457$. In Fig. \ref{figmu} it is
shown the behaviour of $\mu$ for different values of $\eta$, for
$\eta=0$ (dashed line) $\mu=1$ as expected and for $\eta \approx
.274457$ (solid line) it is seen that $\mu$ goes to zero at
$\tau=3\pi/4$.

By direct measuring a quasiprobability distribution, namely the
Wigner function in the lossless case \cite{lut}, i.e. for only one
parameter ($\gamma=0$), or the $Q$-function in our case for a
range of parameters ($\gamma\approx .274457 \chi $), one can
obatin complete information about the state of the field as
quasiprobability distributions contain complete information of the
density matrix \cite{sas}. Moreover the Wigner and $Q$ functions
are related by integral or differential equations, for instance
\cite{kni}
\begin{equation}
W(\alpha) = e^{-\frac{1}{2}\frac{\partial}{\partial \alpha}
\frac{\partial}{\partial \alpha^*}}Q(\alpha).
\end{equation}

In conclusion, we have solved the dispersive interaction between a
quantized elecromagnetic field and a two level atom in the case of
a real cavity (subject to losses) by utilizing superoperator
techniques. We then have used those results to show that even in
the dissipative case we can still obtain information about the
initial cavity field by means of the $Q$-function (unlike the non
dissipative case \cite{lut} where it is reconstructed the Wigner
function). Both functions, being quasiprobability distributions
contain complete information about the state of the cavity field.
In this way, we have been able to extend the range of parameters
in which complete information may be obtained in CQED, from $\gamma=0$
to $\gamma=\approx .274457 \chi$, i.e. by doing a right tuning
complete information of the cavity field may be obtained despite of
cavity losses.

\begin{figure}[hbt]
\caption{\label{figmu} We plot  $\mu$ as a function of $\tau$ for
$\eta=0$ (dashed line), $\eta=0.025$ (dashed-dotted line) and
$\eta \approx .274457$ (solid line)}
\end{figure}

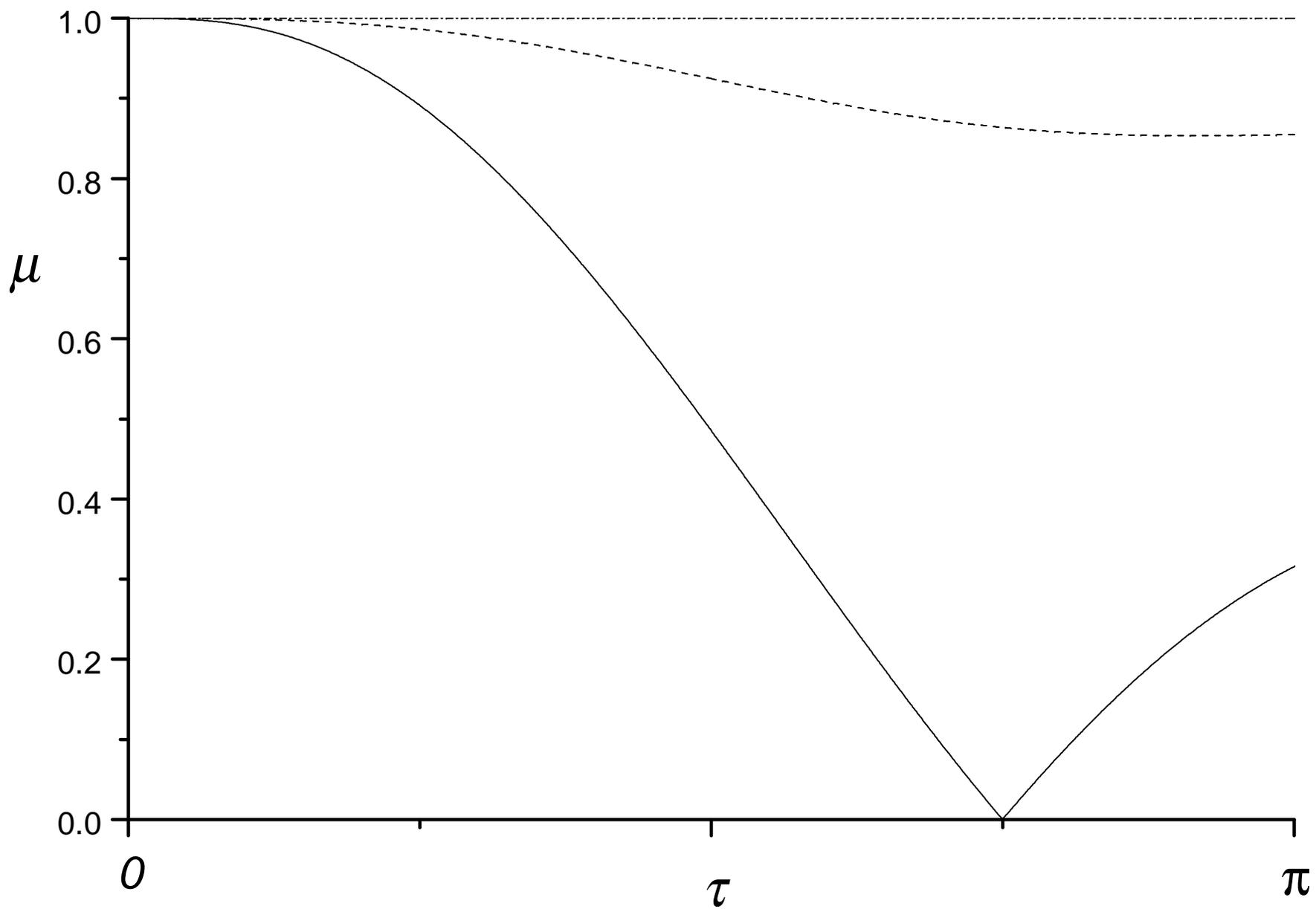
\end{document}